\documentclass[notitlepage,a4paper,aps,prd,onecolumn,superscriptaddress,nofootinbib,groupedaddress,longbibliography]{revtex4-2}
\usepackage{amsmath}
\usepackage{amsfonts}
\usepackage{amssymb}
\usepackage[utf8x]{inputenc}
\usepackage[T1]{fontenc}
\usepackage[colorlinks]{hyperref}
\usepackage{graphicx}

\newcommand{\dd}{\mathrm{d}}

\begin{document}

\title{Kinetic gases in static spherically symmetric modified dispersion relations}

\author{Manuel Hohmann}
\email{manuel.hohmann@ut.ee}
\affiliation{Laboratory of Theoretical Physics, Institute of Physics,  University of Tartu, W. Ostwaldi 1, 50411 Tartu, Estonia}

\begin{abstract}
We study the dynamics of a collisionless kinetic gas in the most general static, spherically symmetric dispersion relation. For a static, spherically symmetric kinetic gas, we derive the most general solution to these dynamics, and find that any solution is given by a one-particle distribution function which depends on three variables. For two particular solutions, describing a shell of monoenergetic orbiting particles and a purely radial inflow, we calculate the particle density as a function of the radial coordinate. As a particular example, we study a $\kappa$-Poincaré modification of the Schwarzschild metric dispersion relation and derive its influence on the particle density. Our results provide a possible route towards quantum gravity phenomenology via the observation of matter dynamics in the vicinity of massive compact objects.
\end{abstract}

\maketitle


\section{Introduction}\label{sec:intro}
Our current understanding of the gravitational interaction, which is most commonly modeled by general relativity (GR), is challenged by a number of open questions: most prominently, these are the unexplained tension of its straightforward cosmological model by observations~\cite{Planck:2018vyg,DiValentino:2021izs} and the lack of a comprehensive quantum theory of gravity, which would allow for a unified quantum description of all fundamental forces. Despite these difficulties, GR has proven to provide a very accurate description of a wide range of observations, ranging from laboratory experiments and the Earth orbit to solar system scales and supermassive black holes. These observations show that any deviation of a gravity theory from GR, which could address the aforementioned open questions, is severely constrained to produce only tiny deviations within those regimes well modeled by GR, while allowing for larger deviations only in regimes which have so far evaded experimental tests~\cite{Baker:2014zba}. Developing and ultimately testing theories which fit into this narrow band between agreement with well-known observations and addressing the questions left open by GR has become a challenging task~\cite{Heisenberg:2018vsk,CANTATA:2021ktz}.

The lack of a complete quantum theory of gravity, and the difficulty of a finding an \emph{ab initio} theory from countless possible modifications of the mathematical description of the gravitational interaction, has motivated the study of \emph{effective} models of quantum gravity, which aim to provide simplified models for particular aspects of the gravitational interaction, without the need to specify the underlying, fundamental theory of gravity, from which these would ultimately have to emerge. Within their regime of validity, such effective models can then be constrained by observations, and thus provide a tool to test also fundamental theories of gravity, by placing an intermediate stepping stone between theory and experiment. A common class of such effective models aims to describe the dynamics of particles and waves by modified dispersion relations (MDR)~\cite{Amelino-Camelia:2008aez,Raetzel:2010je,Addazi:2021xuf}. These models are motivated by the idea that particles propagating through spacetime with larger energy-momentum, and thus characterized by a shorter de Broglie wavelength, would interact more strongly with a possible quantum nature of spacetime, which becomes apparent only on shorter length scales, possibly reaching down to the Planck length. This quantum interaction would become manifest in a dispersion relation which is close to that given by GR at low energies, but exhibits larger deviations at higher energies. Commonly studied example of MDRs include models derived from doubly or deformed special relativity (DSR)~\cite{Amelino-Camelia:2000stu,Amelino-Camelia:2000cpa,Kowalski-Glikman:2002iba}, more generally deformed relativistic kinematics~\cite{Carmona:2019oph,Carmona:2019fwf,Carmona:2012un,Pfeifer:2021tas}, non-commutative spacetime geometries~\cite{Lukierski:1991pn,Majid:1994cy,Snyder:1946qz,Lobo:2016blj}, string theory~\cite{Ellis:1999sd,Amelino-Camelia:1996bln}, loop quantum gravity~\cite{Assanioussi:2014xmz,Amelino-Camelia:2016gfx,Brahma:2018rrg,Lobo:2019jdz}, the Standard-Model Extension (SME)~\cite{Bluhm:2005uj}, as well as the propagation of fields through media~\cite{Klimes:2002,Cerveny:2002,Yajima:2009,Antonelli:2003,Gibbons:2011ib,Rubilar:2002vfs,Punzi:2007di,Perlick:2000}.

Different mathematical descriptions exist for the study of MDRs. One possible description, which is motivated by the general idea to derive the equations of motion for test particles from an action given by a length integral which generalizes the clock postulate from GR, relies on Finsler geometry~\cite{Bao:2000,Bucataru:2007,Girelli:2006fw,Amelino-Camelia:2014rga,Pfeifer:2019wus}. Various proposals exist for a Finsler geometric description of spacetime~\cite{Beem:1970,Pfeifer:2011tk,Minguzzi:2014aua,Javaloyes:2019,Hohmann:2020mgs,Lammerzahl:2012kw,Hasse:2019zqi,Bejancu:2000,Caponio:2015hca,Caponio:2020ofw}, observers~\cite{Hohmann:2013fca,Hohmann:2014gpa,Bernal:2020bul} and the gravitational interaction~\cite{Rutz:1993,Chen:2008,Pfeifer:2011xi,Hohmann:2018rpp}. An alternative, complementary description is given by the Hamiltonian picture~\cite{Miron:2001,Barcaroli:2015xda}, which makes use of a modification of the point particle Hamiltonian obtained from the metric in GR~\cite{Loret:2017lox,Pfeifer:2019zhc,Relancio:2020zok,Relancio:2020rys,Relancio:2022mia}.

Once the dynamics of point particles are known, also the dynamics of ensembles of point particles are straightforward to derive. In the continuum limit, this leads to the model of a kinetic gas~\cite{Ehlers:1971}, which is described by a density function on the phase space of single particles, whose dynamics is derived from the point particle dynamics and a model of their interactions; in the most simple case of vanishing interaction between particles, their dynamics follows from the point particle dynamics alone. Kinetic gases have been studied not only in the metric dispersion relation of GR~\cite{Sarbach:2013fya,Sarbach:2013uba}, while being considered as sources of gravity in the Einstein-Vlasov system~\cite{Andreasson:2011ng}, but also governed by MDRs~\cite{Hohmann:2015ywa,Hohmann:2015duq} and providing their gravitational source~\cite{Hohmann:2019sni,Hohmann:2020yia,Hohmann:2021zbt}.

In this work we study the dynamics of a kinetic gas governed by a general, static, spherically symmetric MDR in the Hamiltonian picture. The aim of this study is to find observational signatures for MDRs in the dynamics of gases in the strong gravitational field in the vicinity of massive bodies, such as neutron stars and black holes, by comparing their dynamics with that given by Schwarzschild spacetime~\cite{Rioseco:2016jwc,Rioseco:2017ahs,Gabarrete:2022aii,Gabarrete:2022fbf}. The basis for this study is the general, static, spherically symmetric point particle Hamiltonian~\cite{Laanemets:2022rmn}. As a particular example, we focus on the $\kappa$-Poincaré MDR~\cite{Barcaroli:2017gvg}. We derive the general dynamics of a collisionless kinetic gas in this background MDR, and focus on two static, spherically symmetric gas configurations, in order to study which modification of their dynamics are introduced by the MDR. Throughout the derivation, we make use of the language of differential forms, as it is most suitable for the Hamiltonian formalism we use here; however, we remark that the main results for the examples are valid also as component expressions.

The article is structured as follows. In section~\ref{sec:coordsym}, we introduce the coordinate system we will be using, and the form of the symmetry generators in these coordinates. We then derive the most general, static, spherically symmetric point particle Hamiltonian, as well as the equations of motion for point particles, in these coordinates in section~\ref{sec:mdr}. These are used to derive the dynamics of a kinetic gas in section~\ref{sec:gas}. We then focus on two examples: in section~\ref{sec:shell}, we study a monoenergetic shell of particles, while in section~\ref{sec:inflow}, we study the case of a static radial flow. We end with a conclusion in section~\ref{sec:conclusion}.

\section{Coordinates and symmetry generators}\label{sec:coordsym}
We start our discussion presented in this article by introducing suitable sets of coordinates. When working with static spherically symmetric spacetimes, it is most conventional to use spherical coordinates \((t, r, \varphi, \vartheta)\) on the spacetime manifold \(M\) in order to express the symmetry generating vector fields in the form
\begin{equation}\label{eq:genvecspher}
X_x = \sin\varphi\partial_{\vartheta} + \frac{\cos\varphi}{\tan\vartheta}\partial_{\varphi}\,, \quad
X_y = -\cos\varphi\partial_{\vartheta} + \frac{\sin\varphi}{\tan\vartheta}\partial_{\varphi}\,, \quad
X_z = -\partial_{\varphi}\,, \quad
X_t = \partial_t\,,
\end{equation}
where the first three vector fields are the generators of rotations, while the last vector field generates time translations. Note that this form of the symmetry generators does not fix the coordinates uniquely. The rotation generators are invariant under any coordinate transformation of the form
\begin{equation}
(t, r) \mapsto (t'(t, r), r'(t, r))\,.
\end{equation}
Under this transformation, \(X_t\) becomes
\begin{equation}
X_t = \partial_t = \frac{\partial t'}{\partial t}\partial_t' + \frac{\partial r'}{\partial t}\partial_r'\,,
\end{equation}
and so it retains its form under coordinate transformations of the form
\begin{equation}\label{eq:coordtrans}
t'(t, r) = t + \tau(r)\,, \quad
r'(t, r) = \rho(r)\,.
\end{equation}
In order to use the Hamilton formalism, we need to introduce suitable coordinates also on the cotangent bundle \(T^*M\). The most straightforward construction to obtain such coordinates from any coordinates \((x^{\mu})\) on \(M\) is to use \emph{induced} or \emph{canonical} coordinates, which represent the position and momentum of a test mass. Denoting a covector at a point with coordinates \((x^{\mu})\) as
\begin{equation}
\bar{x}_{\mu}\,\dd x^{\mu}\,,
\end{equation}
we can use the combined coordinates \((x^{\mu}, \bar{x}_{\mu})\) to denote elements of \(T^*M\). Note that any change of the coordinates on the base manifold also induces a change of coordinates on the cotangent bundle. This holds in particular also for the infinitesimal coordinate changes generated by the symmetry generating vector fields \(X\) given by~\eqref{eq:genvecspher}, if we view these as generators of passive diffeomorphisms, i.e., changes of coordinates. The induced change of coordinates on \(T^*M\) is then generated by a vector field \(\hat{X}\) called the \emph{canonical lift} of \(X\), which can be expressed in induced coordinates as
\begin{equation}
\hat{X} = X^{\mu}\partial_{\mu} - \bar{x}_{\nu}\partial_{\mu}X^{\nu}\bar{\partial}^{\mu}\,,
\end{equation}
where we used the notation
\begin{equation}
\partial_{\mu} = \frac{\partial}{\partial x^{\mu}}\,, \quad
\bar{\partial}^{\mu} = \frac{\partial}{\partial\bar{x}_{\mu}}
\end{equation}
for the coordinate vector fields. For the symmetry generators~\eqref{eq:genvecspher}, we then find the canonical lifts
\begin{subequations}
\begin{align}
\hat{X}_x &= \sin\varphi\partial_{\vartheta} + \frac{\cos\varphi}{\tan\vartheta}\partial_{\varphi} + \frac{\bar{\varphi}\cos\varphi}{\sin^2\vartheta}\bar{\partial}^{\vartheta} - \left(\bar{\vartheta}\cos\varphi - \frac{\bar{\varphi}\sin\varphi}{\tan\vartheta}\right)\bar{\partial}^{\vartheta}\,,\\
\hat{X}_y &= -\cos\varphi\partial_{\vartheta} + \frac{\sin\varphi}{\tan\vartheta}\partial_{\varphi} + \frac{\bar{\varphi}\sin\varphi}{\sin^2\vartheta}\bar{\partial}^{\vartheta} - \left(\bar{\vartheta}\sin\varphi + \frac{\bar{\varphi}\cos\varphi}{\tan\vartheta}\right)\bar{\partial}^{\vartheta}\,,\\
\hat{X}_z &= -\partial_{\varphi}\,,\\
\hat{X}_t &= \partial_t\,,
\end{align}
\end{subequations}
While it is possible to work in induced spherical coordinates, it turns out to be more convenient to introduce a different coordinate system on the cotangent bundle, which we can construct as follows. First, we replace the spatial coordinates by Cartesian coordinates
\begin{equation}
x = r\sin\vartheta\cos\varphi\,, \quad
y = r\sin\vartheta\sin\varphi\,, \quad
z = r\cos\vartheta\,,
\end{equation}
whereby the symmetry generating vector fields become
\begin{equation}
X_x = z\partial_y - y\partial_z\,, \quad
X_y = x\partial_z - z\partial_x\,, \quad
X_z = y\partial_x - x\partial_y\,, \quad
X_t = \partial_t\,,
\end{equation}
and their canonical lifts read
\begin{equation}
\hat{X}_x = z\partial_y - y\partial_z + \bar{z}\partial_{\bar{y}} - \bar{y}\partial_{\bar{z}}\,, \quad
\hat{X}_y = x\partial_z - z\partial_x + \bar{x}\partial_{\bar{z}} - \bar{z}\partial_{\bar{x}}\,, \quad
\hat{X}_z = y\partial_x - x\partial_y + \bar{y}\partial_{\bar{x}} - \bar{x}\partial_{\bar{y}}\,, \quad
\hat{X}_t = \partial_t\,,
\end{equation}
Note that the induced coordinates on the cotangent bundle are related by
\begin{equation}
\bar{x} = \bar{r}\sin\vartheta\cos\varphi + \frac{\bar{\vartheta}}{r}\cos\vartheta\cos\varphi - \frac{\bar{\varphi}}{r\sin\vartheta}\sin\varphi\,, \quad
\bar{y} = \bar{r}\sin\vartheta\sin\varphi + \frac{\bar{\vartheta}}{r}\cos\vartheta\sin\varphi - \frac{\bar{\varphi}}{r\sin\vartheta}\cos\varphi\,, \quad
\bar{z} = \bar{r}\cos\vartheta - \frac{\bar{\vartheta}}{r}\sin\vartheta\,.
\end{equation}
It is useful to denote a few quantities with dedicated symbols. These are the \emph{energy}
\begin{equation}\label{eq:energy}
E = -\bar{t}
\end{equation}
and the \emph{radial momentum}
\begin{equation}\label{eq:radmom}
P = \bar{r} = \frac{x\bar{x} + y\bar{y} + z\bar{z}}{r}\,.
\end{equation}
Further, the components of the \emph{angular momentum} are given by
\begin{equation}\label{eq:angmom}
L_x = y\bar{z} - z\bar{y}\,, \quad
L_y = z\bar{x} - x\bar{z}\,, \quad
L_z = x\bar{y} - y\bar{x}\,.
\end{equation}
Note in particular that the angular momentum is, by construction, perpendicular to a plane spanned by the position \((x, y, z)\) and the linear momentum \((\bar{x}, \bar{y}, \bar{z})\). If the angular momentum is non-vanishing, i.e., if the position and linear momentum are not collinear, we can thus always perform a unique rotation of the coordinate system parameterized by angles \((\Theta, \Phi, \Psi)\), such that the first two align the angular momentum with the polar axis, while the last one brings the position to the null meridian of the rotated coordinate system. Denoting the modulus of the angular momentum by \(L\), we thus have the relations
\begin{equation}\label{eq:position}
x = r(\cos\Theta\cos\Phi\cos\Psi - \sin\Phi\sin\Psi)\,, \quad
y = r(\cos\Theta\sin\Phi\cos\Psi + \cos\Phi\sin\Psi)\,, \quad
z = -r\sin\Theta\cos\Psi
\end{equation}
for the position,
\begin{equation}
L_x = L\sin\Theta\cos\Phi\,, \quad
L_y = L\sin\Theta\sin\Phi\,, \quad
L_z = L\cos\Theta
\end{equation}
for the angular momentum and
\begin{subequations}\label{eq:momentum}
\begin{align}
\bar{x} &= \left(P\cos\Psi - \frac{L}{r}\sin\Psi\right)\cos\Theta\cos\Phi - \left(P\sin\Psi + \frac{L}{r}\cos\Psi\right)\sin\Phi\,,\\
\bar{y} &= \left(P\cos\Psi - \frac{L}{r}\sin\Psi\right)\cos\Theta\sin\Phi + \left(P\sin\Psi + \frac{L}{r}\cos\Psi\right)\cos\Phi\,,\\
\bar{z} &= -\left(P\cos\Psi - \frac{L}{r}\sin\Psi\right)\sin\Theta
\end{align}
\end{subequations}
for the linear momentum. By using the relations~\eqref{eq:energy}, \eqref{eq:position} and~\eqref{eq:momentum}, we can now express the Cartesian induced coordinates \((t, x, y, z, \bar{t}, \bar{x}, \bar{y}, \bar{z})\) in terms of the newly introduced coordinates \((t, r, \Theta, \Phi, \Psi, E, P, L)\) on the cotangent bundle. It then follows that the canonical lifts of the symmetry generating vector fields take the form
\begin{equation}\label{eq:symgen}
\hat{X}_x = \sin\Phi\partial_{\Theta} + \frac{\cos\Phi}{\tan\Theta}\partial_{\Phi} - \frac{\cos\Phi}{\sin\Theta}\partial_{\Psi}\,, \quad
\hat{X}_x = -\cos\Phi\partial_{\Theta} + \frac{\sin\Phi}{\tan\Theta}\partial_{\Phi} - \frac{\sin\Phi}{\sin\Theta}\partial_{\Psi}\,, \quad
\hat{X}_z = -\partial_{\Phi}\,, \quad
\hat{X}_t = \partial_t\,,
\end{equation}
Note that the coordinate transformation becomes singular at \(L = 0\), which is in particular the case for purely radial motion. This means that they cannot be used to model a purely radial flow of a kinetic gas, which we will discuss in section~\ref{sec:inflow}.

\section{Spherically symmetric Hamiltonian}\label{sec:mdr}
We now use the coordinates introduced in the previous section to study the most general static spherically symmetric modified dispersion relation. This can be defined using a Hamiltonian \(H: T^*M \to \mathbb{R}\). The symmetry conditions impose that this is constant along the symmetry vector fields~\eqref{eq:symgen}, and so the corresponding dispersion relation must thus be of the form
\begin{equation}\label{eq:disprel}
-\frac{m^2}{2} = H(r, E, P, L)\,,
\end{equation}
where \(m\) is a constant parameter which is identified with the mass of the particles under consideration. In order to derive the equations of motion for test masses governed by this dispersion relation, we further need the symplectic form, which can be obtained from the symplectic potential. In canonical coordinates, the latter is given by
\begin{equation}
\theta = \bar{x}_{\mu}\dd x^{\mu}\,,
\end{equation}
so that the symplectic form reads
\begin{equation}
\omega = \dd\bar{x}_{\mu} \wedge \dd x^{\mu}\,.
\end{equation}
Note that this form is independent of the choice of the base manifold coordinates \((x^{\mu})\), and depends only on the assumption that the coordinates \((\bar{x}_{\mu})\) are the corresponding induced coordinates. The same holds for the symplectic volume
\begin{equation}
\Sigma = \frac{1}{4!}\omega \wedge \omega \wedge \omega \wedge \omega = \dd x^0 \wedge \dd x^1 \wedge \dd x^2 \wedge \dd x^3 \wedge \dd\bar{x}_0 \wedge \dd\bar{x}_1 \wedge \dd\bar{x}_2 \wedge \dd\bar{x}_3\,.
\end{equation}
Using the rotated coordinates, we can write the symplectic potential as
\begin{equation}
\theta = -E\dd t + P\dd r + L(\cos\Theta\,\dd\Phi + \dd\Psi)\,,
\end{equation}
from which then follows the symplectic form
\begin{equation}
\omega = -\dd E \wedge \dd t + \dd P \wedge \dd r + \dd L \wedge (\cos\Theta\,\dd\Phi + \dd\Psi) - L\sin\Theta\,\dd\Theta \wedge \dd\Phi\,,
\end{equation}
as well as the volume form
\begin{equation}
\Sigma = L\sin\Theta\,\dd t \wedge \dd r \wedge \dd\Theta \wedge \dd\Phi \wedge \dd\Psi \wedge \dd E \wedge \dd P \wedge \dd L\,.
\end{equation}
With the help of the symplectic form, we can calculate the Hamilton vector field \(X_H\), whose integral curves are the solutions to Hamilton's equations of motion, as the unique vector field on \(T^*M\) satisfying
\begin{equation}
\iota_{X_H}\omega = -\dd H\,.
\end{equation}
In canonical coordinates, this is given by
\begin{equation}
X_H = \bar{\partial}^{\mu}H\partial_{\mu} - \partial_{\mu}H\bar{\partial}^{\mu}\,.
\end{equation}
In the coordinates we use here, we find the expression
\begin{equation}\label{eq:hamvfmom}
X_H = -\partial_EH\partial_t + \partial_PH\partial_r - \partial_rH\partial_P + \partial_LH\partial_{\Psi}\,.
\end{equation}
Note that the flow of \(X_H\) preserves \(\omega\) and hence also \(\Sigma\) by construction,
\begin{equation}
\mathcal{L}_{X_H}\omega = 0\,, \quad
\mathcal{L}_{X_H}\Sigma = 0\,.
\end{equation}
It is now easy to see that \(E\), \(L\), \(\Theta\) and \(\Phi\) are constants of motion, since the corresponding components of the Hamilton vector field vanish. Further, also \(H\) itself is a constant of motion, since
\begin{equation}
X_HH = \partial_PH\partial_rH - \partial_rH\partial_PH = 0\,.
\end{equation}
Hence, \(X_H\) is tangent to the mass shell, which is the solution to the dispersion relation~\eqref{eq:disprel}. This motivates the introduction of yet another set of coordinates on \(T^*M\), where one of the coordinates is the value of \(H\), and to replace one of the non-constant coordinates by this new coordinate. Here we choose to replace the radial momentum \(P\). For this purpose, it is necessary to invert the relation between \(H\) and \(P\). Note that this is not globally possible in general, as there will usually be inward and outward momenta which lead to the same value of the Hamiltonian; however, for our purposes it will suffice to assume local invertibility, i.e., the local existence of an inverse
\begin{equation}
P(r, E, L, H)\,.
\end{equation}
Of most importance will be the relation between the partial derivatives of these functions, which is most easily found by calculating their total differentials
\begin{equation}
\dd H = \partial_rH\dd r + \partial_EH\dd E + \partial_PH\dd P + \partial_LH\dd L\,, \quad
\dd P = \partial_rP\dd r + \partial_EP\dd E + \partial_LP\dd L + \partial_HP\dd H\,.
\end{equation}
Substituting either \(\dd H\) or \(\dd P\) with the corresponding other equation, we find that the partial derivatives are related by
\begin{equation}
\partial_rH = -\frac{\partial_rP}{\partial_HP}\,, \quad
\partial_EH = -\frac{\partial_EP}{\partial_HP}\,, \quad
\partial_LH = -\frac{\partial_LP}{\partial_HP}\,, \quad
\partial_PH = \frac{1}{\partial_HP}\,,
\end{equation}
or conversely,
\begin{equation}
\partial_rP = -\frac{\partial_rH}{\partial_PH}\,, \quad
\partial_EP = -\frac{\partial_EH}{\partial_PH}\,, \quad
\partial_LP = -\frac{\partial_LH}{\partial_PH}\,, \quad
\partial_HP = \frac{1}{\partial_PH}\,,
\end{equation}
With the help of these relations, we can write the symplectic form as
\begin{equation}
\omega = -\dd E \wedge \dd t + (\partial_EP\dd E + \partial_LP\dd L + \partial_HP\dd H) \wedge \dd r + \dd L \wedge (\cos\Theta\,\dd\Phi + \dd\Psi) - L\sin\Theta\,\dd\Theta \wedge \dd\Phi\,,
\end{equation}
while the volume becomes
\begin{equation}\label{eq:volume}
\Sigma = -L\sin\Theta\partial_HP\,\dd t \wedge \dd r \wedge \dd\Theta \wedge \dd\Phi \wedge \dd\Psi \wedge \dd E \wedge \dd L \wedge \dd H\,.
\end{equation}
Finally, the Hamilton vector field now reads
\begin{equation}\label{eq:hamvfham}
X_H = -\partial_EH\partial_t + \partial_PH\partial_r + \partial_LH\partial_{\Psi}\,,
\end{equation}
and one sees immediately that it is tangent to the level surfaces of \(H\). Note that the coordinate vector fields \(\partial_t, \partial_r, \partial_{\Psi}\) here are different from the vector fields denoted by the same symbols in the previously related coordinates: while the former are by definition tangent to the level surfaces of \(P\), the newly introduced coordinate vector fields are tangent to the level surfaces of \(H\). In the remainder of this article, we will use the coordinates \((t, r, \Theta, \Phi, \Psi, E, L, H)\), unless stated otherwise.

As a particular example, we consider a $\kappa$-Poincaré modified dispersion relation, which takes the general form~\cite{Barcaroli:2017gvg}
\begin{equation}
H = -\frac{2}{\ell^2}\sinh^2\left(\frac{\ell}{2}Z^{\mu}p_{\mu}\right) + \frac{1}{2}e^{\ell Z^{\mu}p_{\mu}}(g^{\mu\nu} + Z^{\mu}Z^{\nu})p^{\mu}p^{\nu}\,,
\end{equation}
with a unit timelike vector field \(Z^{\mu}\), so that \(Z^{\mu}Z^{\nu}g_{\mu\nu} = -1\), and \(\ell\) is the Planck length, which acts as the quantum gravity scale. In the classical limit \(\ell \to 0\), it reduces to the usual metric dispersion relation
\begin{equation}
\lim_{\ell \to 0}H = \frac{1}{2}g^{\mu\nu}p_{\mu}p_{\nu}\,.
\end{equation}
In the spherically symmetric case, the $\kappa$-Poincaré dispersion relation becomes~\cite{Laanemets:2022rmn}
\begin{equation}\label{eq:kappaspher}
H = -\frac{2}{\ell^2}\sinh^2\left[\frac{\ell}{2}(-cE + dP)\right] + \frac{1}{2}e^{\ell(-cE + dP)}\left[(-a + c^2)E^2 - 2cdEP + (b + d^2)P^2 + \frac{L^2}{r^2}\right]\,,
\end{equation}
with free functions \(a, b, c, d\) of the radial coordinate \(r\) only. In the limit \(\ell \to 0\), this reduces to the metric dispersion relation
\begin{equation}\label{eq:schwmet}
\lim_{\ell \to 0}H = -aE^2 + bP^2 + \frac{L^2}{r^2}\,.
\end{equation}
For simplicity, we will study a modified Schwarzschild geometry, where these functions are given by
\begin{equation}
a^{-1} = b = c^{-2} = 1 - \frac{r_S}{r}\,, \quad
d = 0\,,
\end{equation}
where \(r_S\) is the Schwarzschild radius. Note that any significant modification of the Schwarzschild metric dispersion relation is obtained only if
\begin{equation}
\ell(-cE + dP) \gtrsim 1\,.
\end{equation}
In order to demonstrate the qualitative effect of such a modification on the dynamics of the kinetic gas, we will therefore perform numerical calculations where the particle energy and momentum are chosen in fiducial units such that they will satisfy this order of magnitude, and hence are of the order of the Planck scale.

\section{Spherically symmetric kinetic gas}\label{sec:gas}
We can now use the Hamiltonian dynamics of a single test body, which we derived in the previous section, in order to study the dynamics of a kinetic gas, whose constituent particles follow phase space trajectories which are solutions to Hamilton's equations of motion, unless they undergo collisions with each other. The dynamical variable measuring the phase space density of this gas is the one-particle distribution function \(\phi\), which is defined such that
\begin{equation}\label{eq:1pdf}
N[\sigma] = \int_{\sigma}\phi\Omega
\end{equation}
is the number of particle trajectories passing through \(\sigma\) for any hypersurface \(\sigma \subset T^*M\), where
\begin{equation}
\Omega = \iota_{X_H}\Sigma = L\sin\Theta\left(\dd t \wedge \dd\Psi - \partial_EP\,\dd r \wedge \dd\Psi + \partial_LP\,\dd t \wedge \dd r\right) \wedge \dd\Theta \wedge \dd\Phi \wedge \dd E \wedge \dd L \wedge \dd H
\end{equation}
is known as the \emph{particle measure}, which satisfies
\begin{equation}
\mathcal{L}_{X_H}\Omega = 0\,.
\end{equation}
For simplicity, we will restrict ourselves to the case of a collisionless gas, which means that the motion of its constituent particles is governed exclusively by Hamilton's equations of motion. It then follows that the one-particle distribution function satisfies the Liouville equation
\begin{equation}
\mathcal{L}_{X_H}\phi = -\partial_EH\partial_t\phi + \partial_PH\partial_r\phi + \partial_LH\partial_{\Psi}\phi = 0\,,
\end{equation}
which means that \(\phi\) is constant along the integral curves of \(X_H\), i.e., along the phase space trajectories of the constituent particles. In the following, again for simplicity, we restrict ourselves to one-particle distribution functions which share the assumed symmetries of the Hamiltonian,
\begin{equation}\label{eq:sym1pdf}
\hat{X}_x\phi = \hat{X}_y\phi = \hat{X}_z\phi = \hat{X}_t\phi = 0\,,
\end{equation}
and are thus in particular spherically symmetric and static. While this excludes several physically interesting examples, such as accretion discs and time-dependent flows, it will allow us to give explicit formulas for a number of physically relevant quantities, as we will see below. Under the assumed symmetries, the one-particle distribution function is of the form
\begin{equation}
\phi(r, E, L, H)\,.
\end{equation}
In this case the Liouville equation simply becomes
\begin{equation}\label{eq:liouville}
\partial_r\phi = 0\,,
\end{equation}
which means that the one-particle distribution function, in fact, must be of the form
\begin{equation}\label{eq:cl1pdf}
\phi(E, L, H)\,.
\end{equation}
Note that this solution has also been found in the particular case of a Schwarzschild metric dispersion relation~\cite{Rioseco:2016jwc}. It is important to realize that this does \emph{not} mean that it is constant over all of spacetime, due to the apparent absence of any spatial coordinates in the form given above. However, this dependence is encoded implicitly by the fact that we chose the value \(H\) of the Hamiltonian as one of the coordinates, which in turn depends on the radial coordinate \(r\). Indeed, if we use the radial momentum \(P\) as a coordinate instead of \(H\), we see from the expression~\eqref{eq:hamvfmom} of the Hamilton vector field that the Liouville equation simply relates the dependence of \(\phi\) on \(r\) and \(P\) such that these are exactly identical to that of \(H\), and can thus be expressed by having \(\phi\) depend on \(H\) only instead of \(r\) and \(P\) individually.

Note that the one-particle distribution function~\eqref{eq:cl1pdf} depends on constants of motion only, which means that it must be constant along the level sets of these constants. More precisely, it must be constant along the \emph{connected components} of the level sets, since the Liouville equation~\eqref{eq:liouville} and the symmetry conditions~\eqref{eq:sym1pdf} are purely local conditions, which do not pose any restriction on the relation between its values on different connected components. This is relevant as there might be several connected components, for example for incoming and outgoing particles, which have the same values of \(E, L, H\). Since any particle trajectory is confined to a single connected component of a level set, and particles do not interact with each other by the assumption of a collisionless gas, we may thus consider each connected component of a level set individually, and in particular study one-particle distribution functions which are confined to a single connected component. Special care must be taken, since \(\phi\) must be understood as a Dirac-like distribution in the integral~\eqref{eq:1pdf} in this case, which is defined only under taking integrals. This can be taken into account by realizing that every connected component of the aforementioned level sets, which forms a submanifold \(\Upsilon\), is equipped with a unique (up to a constant factor) volume form
\begin{equation}
\Sigma_{\Upsilon} = -L\sin\Theta\partial_HP\,\dd t \wedge \dd r \wedge \dd\Theta \wedge \dd\Phi \wedge \dd\Psi
\end{equation}
which is invariant under the symmetry generators~\eqref{eq:symgen} and the Hamilton vector field~\eqref{eq:hamvfham}, and which gives rise to a particle measure
\begin{equation}
\Omega_{\Upsilon} = \iota_{X_H}\Sigma_{\Upsilon} = L\sin\Theta(\dd t \wedge \dd\Psi - \partial_EP\,\dd r \wedge \dd\Psi + \partial_LP\,\dd t \wedge \dd r) \wedge \dd\Theta \wedge \dd\Phi\,.
\end{equation}
For any hypersurface \(\sigma \in T^*M\) transverse to \(\Upsilon\), one finds that the number of particle trajectories passing \(\sigma\) takes the form
\begin{equation}\label{eq:const1pdf}
N[\sigma] = N[\sigma \cap \Upsilon] = C\int_{\sigma \cap \Upsilon}\Omega_{\Upsilon}
\end{equation}
with an overall normalization constant \(C\) which determines the total number of particles. In the following, we will therefore study the properties of distribution functions of this type for a few example configurations.

\section{Monoenergetic orbiting shell}\label{sec:shell}
As a first example, we will study a static, spherically symmetric shell of particles which are orbiting the spherically symmetric dispersion relation on bound orbits. For this purpose we will assume that the Hamiltonian \(H\) is chosen such that for fixed values of the energy \(E\), angular momentum \(L\) and the Hamiltonian \(H\) itself, there exist two radii \(r_1, r_2\) at which \(\partial_PH = 0\), so that these radii mark the extremal values of the radial coordinate \(r\) along the orbit. Further, we assume that for \(r_1 < r < r_2\) there exist two solutions for \(P\) to the dispersion relation such that \(\partial_PH\) has opposite signs for both solutions, so that particle trajectories oscillate between these values. It follows that the particle density will be non-vanishing only for \(r_1 \leq r \leq r_2\).

\begin{figure}
\includegraphics[width=0.75\textwidth]{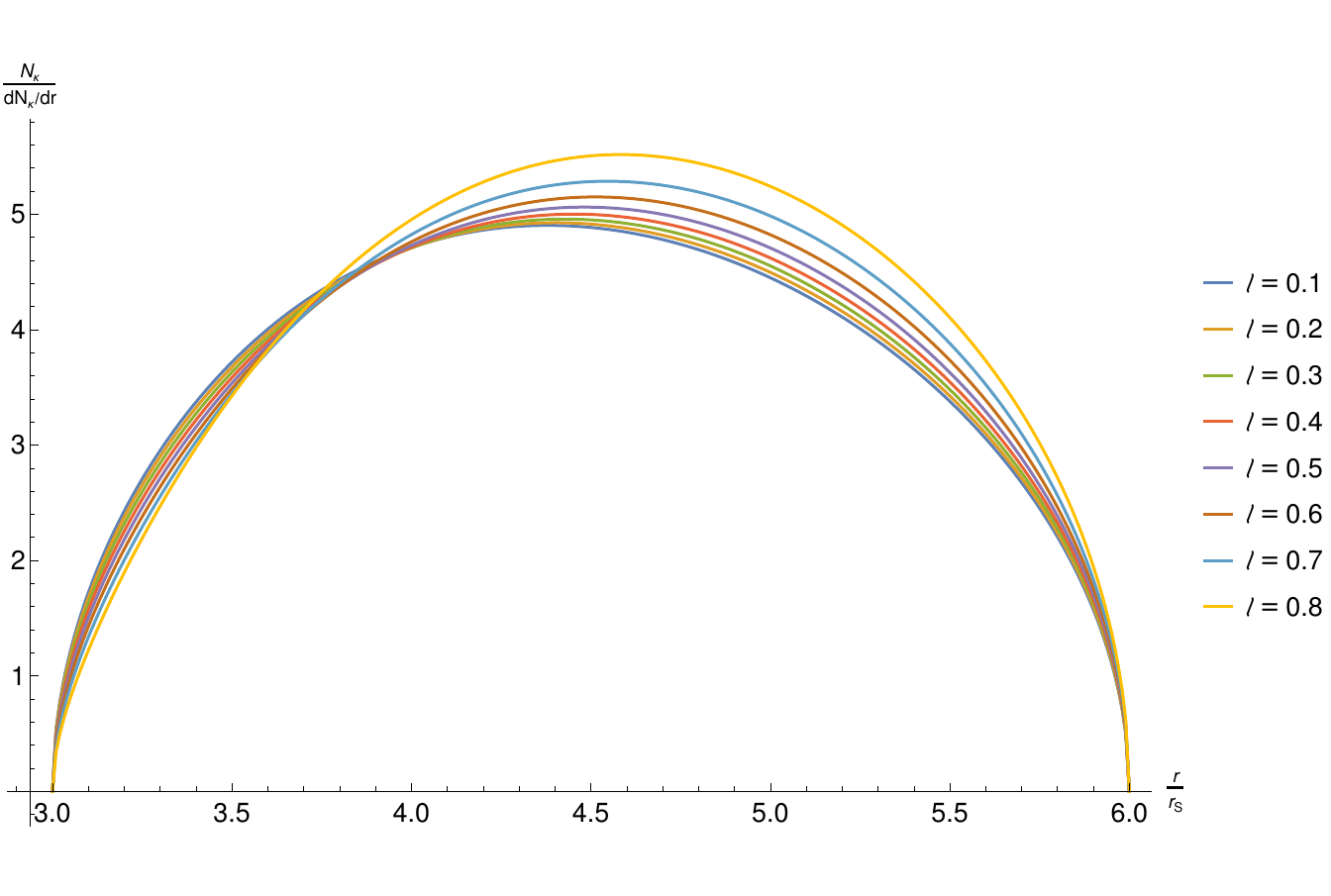}
\caption{Inverse normalized linear density of orbiting kinetic gas for the $\kappa$-Poincaré dispersion relation.}
\label{fig:orbabs}
\end{figure}

\begin{figure}
\includegraphics[width=0.75\textwidth]{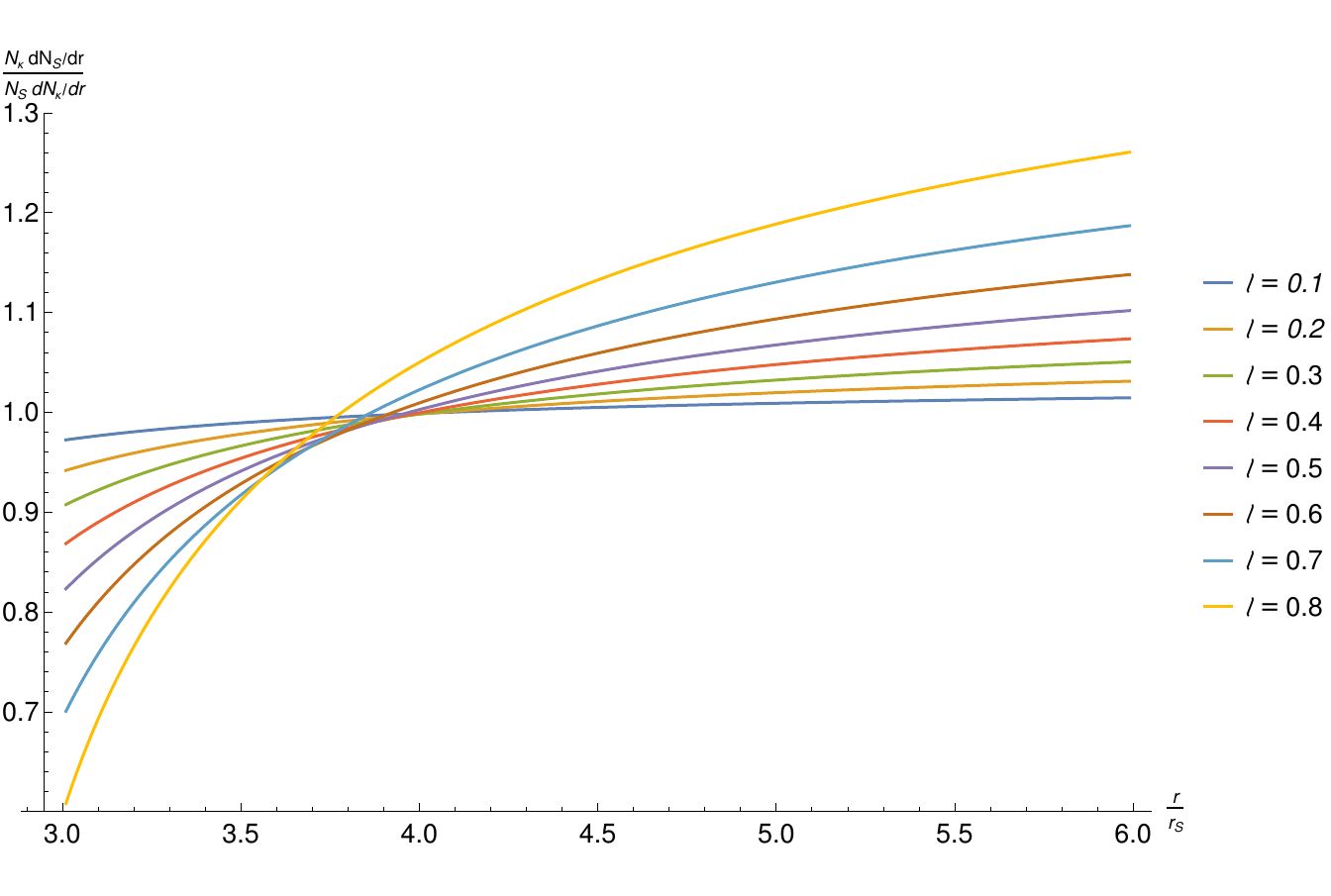}
\caption{Comparison of the linear density of orbiting kinetic gases in $\kappa$-Poincaré and Schwarzschild metric dispersion relations.}
\label{fig:orbrel}
\end{figure}

It follows from the aforementioned assumptions that there are different values for \(P\) for the incoming and outgoing parts of the trajectories, and we will denote these two values by \(P_{\pm}\), where the upper sign denotes the outgoing particles. The phase space density is then given by
\begin{equation}
C_{\pm}\sin\Theta(\dd t \wedge \dd\Psi - \partial_EP_{\pm}\,\dd r \wedge \dd\Psi + \partial_LP_{\pm}\,\dd t \wedge \dd r) \wedge \dd\Theta \wedge \dd\Phi\,,
\end{equation}
with constants \(C_{\pm}\), which determine the phase space densities of incoming and outgoing particles. Note that the Liouville equation must be satisfied also at the turnaround radii \(r_{1,2}\), where the coordinate transformation from \(P\) to \(H\) becomes singular. For the static shell we consider here, this is the case if at each constant value \(r = R\) of the radial coordinate the number of incoming and outgoing particles within the interval \(T_1 < t < T_2\) is equal, so that one must have
\begin{equation}
0 = N[\sigma_R] = (C_+ + C_-)\int_{T_1}^{T_2}\dd t\int_0^{\pi}\sin\Theta\dd\Theta\int_0^{2\pi}\dd\Phi\int_0^{2\pi}\dd\Psi = 8\pi^2(C_+ + C_-)(T_2 - T_1)\,,
\end{equation}
and thus
\begin{equation}
C_+ = -C_- = C\,,
\end{equation}
reflecting the fact that particles passing this hypersurface in opposite directions are counted with opposite signs. The number of incoming and outgoing particle trajectories passing through a hypersurface \(\sigma_T\) of constant time \(t = T\) bounded by \(R_1 < r < R_2\) is then given by
\begin{equation}
N_{\pm}[\sigma_T] = -C_{\pm}\int_{R_1}^{R_2}\partial_EP_{\pm}\dd r\int_0^{\pi}\sin\Theta\dd\Theta\int_0^{2\pi}\dd\Phi\int_0^{2\pi}\dd\Psi = -8\pi^2C_{\pm}\int_{R_1}^{R_2}\partial_EP_{\pm}\dd r\,,
\end{equation}
and so the total number of particle trajectories passing through this hypersurface is
\begin{equation}
N[\sigma_T] = N_+[\sigma_T] + N_-[\sigma_T] = -8\pi^2C\int_{R_1}^{R_2}(\partial_EP_+ - \partial_EP_-)\dd r\,.
\end{equation}
Instead of this integral expression, we can write the linear particle density
\begin{equation}\label{eq:lindensorb}
\frac{\dd N[\sigma_T]}{\dd r} = -8\pi^2C(\partial_EP_+ - \partial_EP_-)\,.
\end{equation}
As a particular example, we apply our findings to the $\kappa$-Poincaré dispersion relation~\eqref{eq:kappaspher}, and we will fix the energy \(E\), angular momentum \(L\) and Hamiltonian \(H\) such that the turnaround radii are \(r_1 = 3r_S\) and \(r_2 = 6r_S\). Note that we cannot solve for these parameter values analytically, except in the Schwarzschild case \(\ell \to 0\). For the latter, we can choose
\begin{equation}
E = \frac{\sqrt{10}}{6}\,, \quad
\frac{L}{r_S} = 1\,, \quad
H = -\frac{11}{72}
\end{equation}
in fiducial units, in order to achieve the desired turnaround radii. For \(\ell > 0\), we will keep \(H\) fixed, and numerically solve for \(E\) and \(L\), such that the turnaround radii remain fixed. We then need to determine the momenta \(P_{\pm}\) and their derivatives \(\partial_EP_{\pm}\) along the particle trajectories. Also these can be obtained analytically only in the Schwarzschild case \(\ell \to 0\), where we have
\begin{equation}
\lim_{\ell \to 0}P_{\pm}(r, E, L, H) = \pm\sqrt{\frac{r}{r - r_S}\left(2H + \frac{rE^2}{r - r_S} - \frac{L^2}{r^2}\right)}
\end{equation}
and
\begin{equation}
\lim_{\ell \to 0}\partial_EP_{\pm}(r, E, L, H) = \frac{r^2E}{(r - r_S)^2P_{\pm}}\,,
\end{equation}
and must be determined numerically for \(\ell > 0\). Note in particular that at the turnaround radii \(r = r_{1,2}\) one has \(P_{\pm} = 0\), so that \(\partial_EP_{\pm}\) and thus also the linear particle density diverge. Nevertheless, the total number
\begin{equation}
N[\sigma_T] = -8\pi^2C\int_{r_1}^{r_2}(\partial_EP_+ - \partial_EP_-)\dd r
\end{equation}
of particle trajectories is still finite. To eliminate the constant \(C\), it is therefore most convenient to normalize the linear particle density by the total particle number, and plot its inverse
\begin{equation}
N[\sigma_T]\left(\frac{\dd N[\sigma_T]}{\dd r}\right)^{-1}\,,
\end{equation}
which vanishes at the turnaround radii, for different values of \(\ell\). The result is shown for the $\kappa$-Poincaré dispersion relation \(N = N_{\kappa}\) in figure~\ref{fig:orbabs}. We see that for larger values of \(\ell\) the density increases at smaller radii, while it decreases at larger radii. To see this effect more clearly also at the turnaround radii, we calculate the relative change of the density compared to the Schwarzschild case \(N = N_S\). This is shown in figure~\ref{fig:orbrel}.

\section{Monoenergetic radial inflow}\label{sec:inflow}
\begin{figure}
\includegraphics[width=0.75\textwidth]{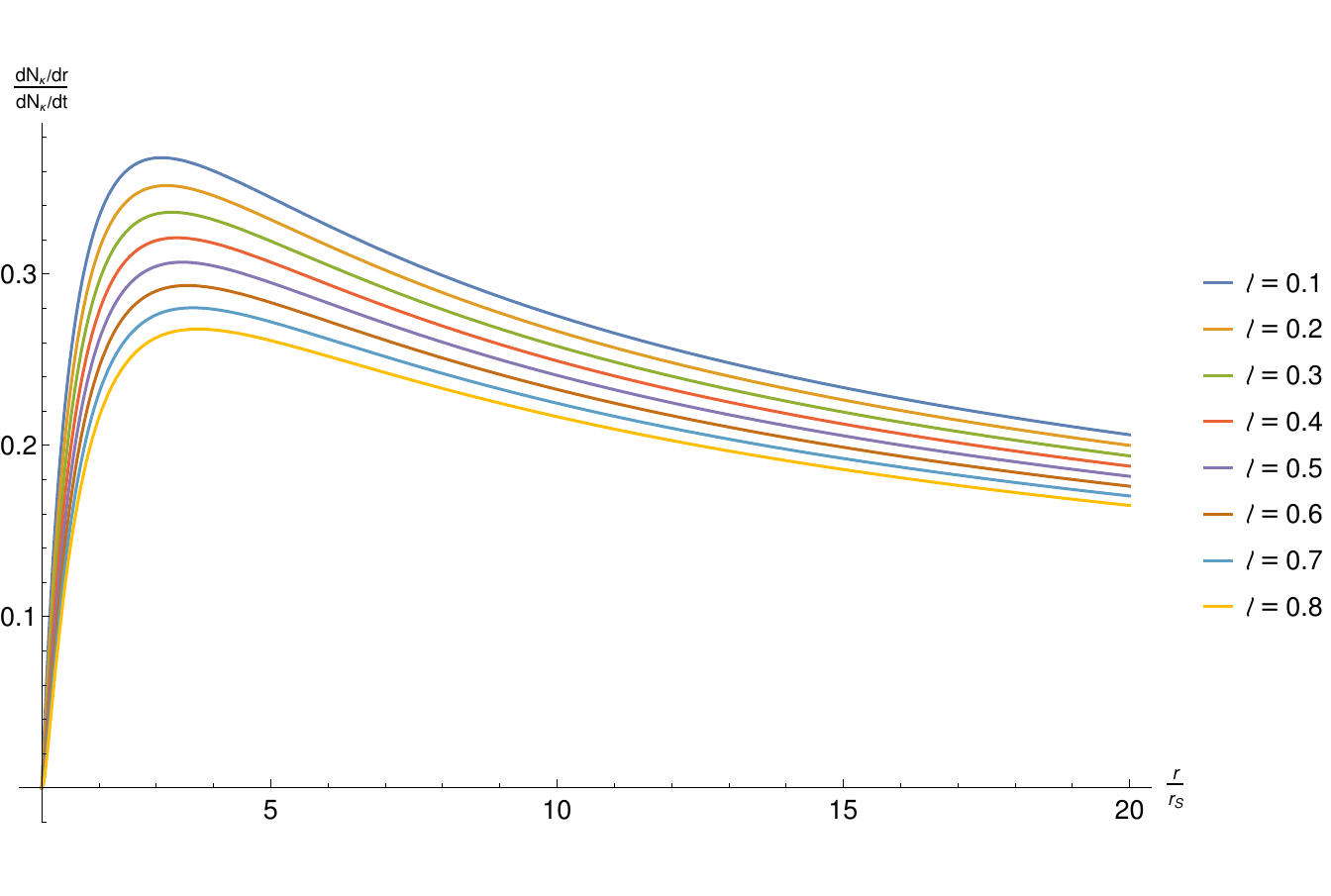}
\caption{Linear density of radially inflowing, marginally bound kinetic gas for the $\kappa$-Poincaré dispersion relation.}
\label{fig:flowlin}
\end{figure}

\begin{figure}
\includegraphics[width=0.75\textwidth]{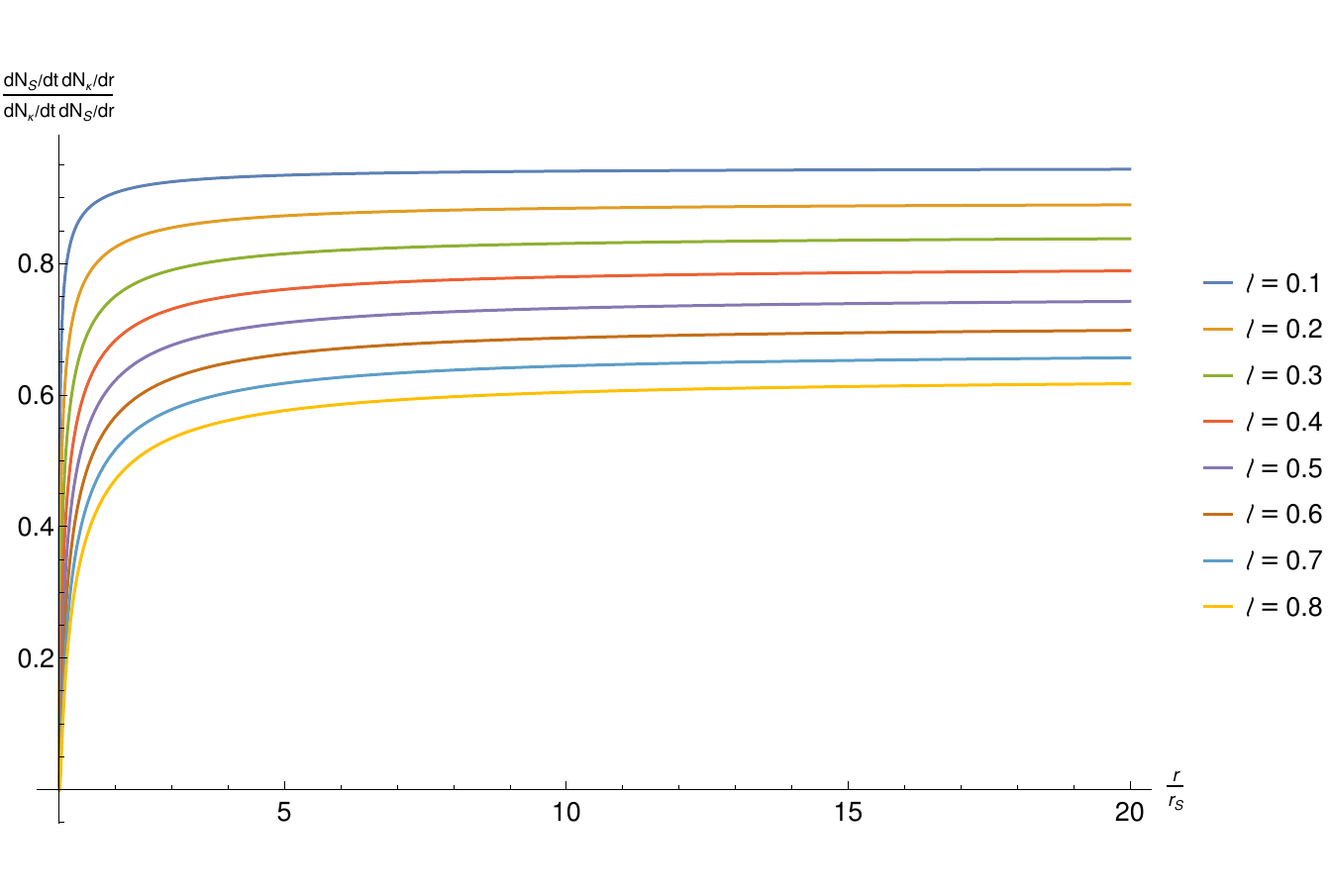}
\caption{Comparison of the linear density of radially inflowing kinetic gases in $\kappa$-Poincaré and Schwarzschild metric dispersion relations.}
\label{fig:flowrel}
\end{figure}

We now study the case of a spherically symmetric, static kinetic gas composed of particles with constant energy and mass which flows radially towards the center of the spherically symmetric dispersion relation. As argued in section~\ref{sec:coordsym}, we cannot use the coordinates which we constructed from aligning the coordinate system with the angular momentum in this case, since the latter vanishes, and so the coordinate transformation becomes singular. Hence, we will proceed in spherical spacetime coordinates and the corresponding induced coordinates on the cotangent bundle, which turn out to be suitable in this case. We can use these to write the angular momentum as
\begin{equation}
L = \sqrt{\bar{\vartheta}^2 + \frac{\bar{\varphi}^2}{\sin^2\vartheta}}\,.
\end{equation}
Since we are working in induced coordinates, the symplectic potential takes the canonical form
\begin{equation}
\theta = \bar{t}\,\dd t + \bar{r}\,\dd r + \bar{\vartheta}\,\dd\vartheta + \bar{\varphi}\,\dd\varphi\,,
\end{equation}
while the symplectic form reads
\begin{equation}
\omega = \dd\bar{t} \wedge \dd t + \dd\bar{r} \wedge \dd r + \dd\bar{\vartheta} \wedge \dd\vartheta + \dd\bar{\varphi} \wedge \dd\varphi\,.
\end{equation}
The symplectic volume thus becomes
\begin{equation}
\Sigma = \dd t \wedge \dd r \wedge \dd\vartheta \wedge \dd\varphi \wedge \dd\bar{t} \wedge \dd\bar{r} \wedge \dd\bar{\vartheta} \wedge \dd\bar{\varphi}\,.
\end{equation}
With the help of these quantities, we can calculate the Hamilton vector field
\begin{equation}
X_H = -\partial_EH\partial_t + \partial_PH\partial_r - \partial_rH\bar{\partial}^r + \frac{\partial_LH}{\sqrt{\bar{\vartheta}^2 + \frac{\bar{\varphi}^2}{\sin^2\vartheta}}}\left(\bar{\vartheta}\partial_{\vartheta} + \frac{\bar{\varphi}}{\sin^2\vartheta}\partial_{\varphi} + \frac{\bar{\varphi}^2}{\sin^2\vartheta\tan\vartheta}\bar{\partial}^{\vartheta}\right)\,,
\end{equation}
from which further follows the particle measure
\begin{equation}
\begin{split}
\Omega &= (\partial_EH\dd r \wedge \dd\bar{r} + \partial_PH\dd t \wedge \dd\bar{r} + \partial_rH\dd t \wedge \dd r) \wedge \dd\vartheta \wedge \dd\varphi \wedge \dd\bar{t} \wedge \dd\bar{\vartheta} \wedge \dd\bar{\varphi}\\
&\phantom{=}+ \frac{\partial_LH}{\sqrt{\bar{\vartheta}^2 + \frac{\bar{\varphi}^2}{\sin^2\vartheta}}}\left(\bar{\vartheta}\dd\varphi \wedge \dd\bar{\vartheta} - \frac{\bar{\varphi}}{\sin^2\vartheta}\dd\vartheta \wedge \dd\bar{\vartheta} + \frac{\bar{\varphi}^2}{\sin^2\vartheta\tan\vartheta}\dd\vartheta \wedge \dd\varphi\right) \wedge \dd\bar{\varphi} \wedge \dd t \wedge \dd r \wedge \dd\bar{t} \wedge \dd\bar{r}\,.
\end{split}
\end{equation}
In order to explicitly integrate out the mass shell, we introduce again coordinates adapted to the latter, by replacing the radial momentum \(\bar{r}\) by the value \(H\) of the Hamiltonian. This is most straightforward for the differential forms, using the relation
\begin{equation}
\dd\bar{r} = \partial_rP\dd r - \partial_EP\dd\bar{t} + \partial_HP\dd H + \frac{\partial_LP}{\sqrt{\bar{\vartheta}^2 + \frac{\bar{\varphi}^2}{\sin^2\vartheta}}}\left(\bar{\vartheta}\dd\bar{\vartheta} + \frac{\bar{\varphi}}{\sin^2\vartheta}\dd\bar{\varphi} - \frac{\bar{\varphi}^2}{\sin^2\vartheta\tan\vartheta}\dd\vartheta\right)\,.
\end{equation}
In particular, we have the symplectic form
\begin{equation}
\omega = \dd\bar{t} \wedge \dd t + \left[-\partial_EP\dd\bar{t} + \partial_HP\dd H + \frac{\partial_LP}{\sqrt{\bar{\vartheta}^2 + \frac{\bar{\varphi}^2}{\sin^2\vartheta}}}\left(\bar{\vartheta}\dd\bar{\vartheta} + \frac{\bar{\varphi}}{\sin^2\vartheta}\dd\bar{\varphi} - \frac{\bar{\varphi}^2}{\sin^2\vartheta\tan\vartheta}\dd\vartheta\right)\right] \wedge \dd r + \dd\bar{\vartheta} \wedge \dd\vartheta + \dd\bar{\varphi} \wedge \dd\varphi\,.
\end{equation}
and its volume form
\begin{equation}
\Sigma = \partial_HP\dd t \wedge \dd r \wedge \dd\vartheta \wedge \dd\varphi \wedge \dd\bar{t} \wedge \dd H \wedge \dd\bar{\vartheta} \wedge \dd\bar{\varphi}\,.
\end{equation}
From these one obtains the Hamilton vector field
\begin{equation}
X_H = -\partial_EH\partial_t + \partial_PH\partial_r + \frac{\partial_LH}{\sqrt{\bar{\vartheta}^2 + \frac{\bar{\varphi}^2}{\sin^2\vartheta}}}\left(\bar{\vartheta}\partial_{\vartheta} + \frac{\bar{\varphi}}{\sin^2\vartheta}\partial_{\varphi} + \frac{\bar{\varphi}^2}{\sin^2\vartheta\tan\vartheta}\bar{\partial}^{\vartheta}\right)\,,
\end{equation}
and finally the particle measure
\begin{equation}
\begin{split}
\Omega &= (\dd t \wedge \dd H - \partial_EP\dd r \wedge \dd H) \wedge \dd\vartheta \wedge \dd\varphi \wedge \dd\bar{t} \wedge \dd\bar{\vartheta} \wedge \dd\bar{\varphi}\\
&\phantom{=}- \frac{\partial_LP}{\sqrt{\bar{\vartheta}^2 + \frac{\bar{\varphi}^2}{\sin^2\vartheta}}}\left(\bar{\vartheta}\dd\varphi \wedge \dd\bar{\vartheta} - \frac{\bar{\varphi}}{\sin^2\vartheta}\dd\vartheta \wedge \dd\bar{\vartheta} + \frac{\bar{\varphi}^2}{\sin^2\vartheta\tan\vartheta}\dd\vartheta \wedge \dd\varphi\right) \wedge \dd\bar{\varphi} \wedge \dd t \wedge \dd r \wedge \dd\bar{t} \wedge \dd H\,.
\end{split}
\end{equation}
We can now study the restriction to a submanifold \(\Upsilon\) with fixed values of the energy \(E = -\bar{t}\), Hamiltonian \(H\) and purely radial flow \(L = 0\), so that \(\bar{\vartheta} = \bar{\varphi} = 0\). For this purpose we will assume that the Hamiltonian \(H\), and hence also the radial momentum \(P\) obtained from the latter, satisfies the regularity conditions
\begin{equation}
\partial_LH|_{L = 0} = \partial_LP|_{L = 0} = 0\,,
\end{equation}
so that it is differentiable with respect to the momentum components \(\bar{\vartheta}\) and \(\bar{\varphi}\). It then follows that the angular components of \(X_H\) and \(\Omega\) vanish on \(\Upsilon\). By integrating out any coordinates transverse to \(\Upsilon\), we are left with the reduced particle measure
\begin{equation}
\Omega_{\Upsilon} = \sin\vartheta(\dd t - \partial_EP\dd r) \wedge \dd\vartheta \wedge \dd\varphi\,,
\end{equation}
which is invariant under the flow of the symmetry generators~\eqref{eq:symgen} and the restricted Hamilton vector field
\begin{equation}
X_H|_{\Upsilon} = -\partial_EH\partial_t + \partial_PH\partial_r\,.
\end{equation}
Using the fact that the restriction of the one-particle distribution function to \(\Upsilon\) is given by a constant \(C\), we can then calculate the number of particle trajectories through any hypersurface \(\sigma\) from the integral~\eqref{eq:const1pdf}. For a hypersurface \(\sigma_R\) with constant radius \(r = R\) and time range \(T_1 < t < T_2\) we have
\begin{equation}
N[\sigma_R] = C\int_{T_1}^{T_2}\dd t\int_0^{\pi}\sin\vartheta\dd\vartheta\int_0^{2\pi}\dd\varphi = 4\pi C(T_2 - T_1)\,,
\end{equation}
which is independent of the radius \(R\), since the same number of particle trajectories must pass through any hypersurface of this type, and the flow rate
\begin{equation}\label{eq:flowrate}
\frac{\dd N[\sigma_R]}{\dd t} = 4\pi C
\end{equation}
is constant. Considering a hypersurface \(\sigma_T\) at constant \(t = T\) with \(R_1 < r < R_2\) we have
\begin{equation}
N[\sigma_T] = -C\int_{R_1}^{R_2}\partial_EP\dd r\int_0^{\pi}\sin\vartheta\dd\vartheta\int_0^{2\pi}\dd\varphi = -4\pi C\int_{R_1}^{R_2}\partial_EP\dd r\,,
\end{equation}
and so the linear particle density is given by
\begin{equation}\label{eq:lindensrad}
\frac{\dd N[\sigma_T]}{\dd r} = -4\pi C\partial_EP\,.
\end{equation}
To illustrate our result, we consider once again the $\kappa$-Poincaré dispersion relation~\eqref{eq:kappaspher}, where we now set \(L = 0\) to model the radial flow. Further, we restrict ourselves to the marginally bound case
\begin{equation}
\lim_{r \to \infty}\partial_PH = 0\,.
\end{equation}
Keeping the value of \(H\) identical to the previous section, we can achieve this for the Schwarzschild case \(\ell \to 0\) by
\begin{equation}
E = \frac{\sqrt{11}}{6}\,, \quad
H = -\frac{11}{72}
\end{equation}
in fiducial units, while resorting to a numerical solution for \(E\) in the general case \(\ell > 0\). In order to eliminate the normalization constant \(C\), we cannot use the total number of particle trajectories in this case, since the latter is infinite, and so we will normalize the particle density by the flow rate~\eqref{eq:flowrate}. Also it turns out to be more convenient to plot the linear particle density itself, instead of its inverse as done in the previous section, since the latter turns out to diverge. Hence, we will plot the expression
\begin{equation}
\frac{\dd N[\sigma_T]}{\dd r}\left(\frac{\dd N[\sigma_R]}{\dd t}\right)^{-1}\,.
\end{equation}
This is shown for the $\kappa$-Poincaré dispersion relation \(N = N_{\kappa}\) in figure~\ref{fig:flowlin}. We see that for larger values of \(\ell\), a lower particle density leads to the same flow rate, which means that the marginally bound particles propagate at the higher coordinate velocity \(-\dd r/\dd t\) in this case. We compare this with the result obtained in the Schwarzschild case \(N = N_S\), and display the ratio of the densities for a fixed flow rate in figure~\ref{fig:flowrel}. Most remarkably, we find that at the horizon \(r = r_S\) this ratio becomes zero for any non-zero value of the Planck length, which indicates its dominating effect on the ultrarelativistic flow near the horizon.

\section{Conclusion}\label{sec:conclusion}
We have studied the dynamics of a collisionless kinetic gas governed by a static, spherically symmetric dispersion relation. By introducing a suitable set of coordinates, we have derived the general form of the Liouville equation from Hamilton's equations of motion for point particles. Under the assumption of a static, spherically symmetric kinetic gas, we have derived the most general solution to the Liouville equation, and found that it is given by a one-particle distribution function which depends on three quantities, which are the energy \(E\), angular momentum \(L\) and Hamiltonian \(H\). Using this result, we have studied two example configurations for which the one-particle distribution function is described by a Dirac-like distribution in these variables, being non-zero only on a single level set. These configurations are given by a monoenergetic shell of orbiting particles and by a monoenergetic radial inflow. For both configurations we have calculated the linear particle density \(\dd N / \dd r\) as a function of the radial coordinate. As a particular example, we have considered a $\kappa$-Poincaré modification of the Schwarzschild metric dispersion relation, and numerically determined the linear particle density. For the orbiting shell, we have found that in the $\kappa$-Poincaré dispersion relation the particle density is shifted towards lower radii compared to the Schwarzschild metric distribution function. For the radial inflow, we have found that while keeping a constant flow rate, the particle density in the $\kappa$-Poincaré case is lower than in the Schwarzschild case. Note that while the derivation is based on the language of differential forms, the resulting expressions for the linear particle density are the simple component expressions~\eqref{eq:lindensorb} and~\eqref{eq:lindensrad}.

Our results are twofold. As a general result, we have put forward the study of modified dispersion relations by their effects on a kinetic gas under the assumption of spherical symmetry and symmetry under time translation. Applying this result to further modified dispersion relations, in particular those derived from quantum gravity theories mentioned in the introduction, or other modifications of the Schwarzschild metric~\cite{Kapsabelis:2022bue,Cheraghchi:2022zgv,Voicu:2023fey}, establishes a window towards quantum gravity phenomenology through the observation of gas dynamics in the vicinity of massive, compact objects, such as neutron stars and black holes. As a particular result, we have found these effects for a $\kappa$-Poincaré modified dispersion relation. The latter can easily be generalized to other dispersion relations by making use of our general result, and we aim to do so in future research. Another possible line of future investigations is obtained by relaxing the symmetry conditions on the kinetic gas, in order to study the dynamics of accretion disks, jets and tidal disruption events. Relaxing also the condition of spherical symmetry on the modified dispersion relation, a generalization to rotating black holes may be studied.

\begin{acknowledgments}
MH gratefully acknowledges the full financial support by the Estonian Research Council through the Personal Research Funding project PRG356.
\end{acknowledgments}

\bibliography{sphergas}
\end{document}